\documentclass[twocolumn,preprintnumbers,amsmath,amssymb]{revtex4-1}
\usepackage{graphicx}
\usepackage{dcolumn}
\usepackage{bm}
\usepackage[]{natbib}
\usepackage{hyperref}

\preprint{Submitted to Phys. Rev. C (October 2019)}

\begin{document}
\title{Fusion of Borromean nucleus $^{9}$Be with $^{197}$Au target at near barrier energies}

\author{Malika Kaushik$^{1}$}
\author{G.~Gupta$^{2}$}
\author{Swati Thakur$^{1}$}
 \author{H.~Krishnamoorthy$^{3,4}$}
 \author{Pushpendra~P.~Singh$^{1}$}
 \author{V.V.~Parkar$^{5}$}
 \author{V.~Nanal$^{2}$}
  \email{nanal@tifr.res.in}
\author{A.~Shrivastava$^{4,5}$}
\author{R.G.~Pillay$^{1}$}
\author{K.~Mahata$^{4,5}$}
\author{K.~Ramachandran$^{5}$}
\author{S.~Pal$^{6}$}
\author{C.S.~Palshetkar$^{2}$}
\author{S.K.~Pandit$^{5}$}
\affiliation{$^{1}$Department of Physics, Indian Institute of Technology Ropar, Rupnagar - 140 001, Punjab, India}
\affiliation{$^{2}$Department of Nuclear and Atomic Physics, Tata Institute of Fundamental Research, Mumbai - 400 005, India}
\affiliation{$^{3}$India-based Neutrino Observatory, Tata Institute of Fundamental Research, Mumbai - 400 005, India}
\affiliation{$^{4}$Homi Bhabha National Institute, Anushaktinagar, Mumbai - 400094, India}
\affiliation{$^{5}$Nuclear Physics Divison, Bhabha Atomic Research Centre, Mumbai - 400085, India}
\affiliation{$^{6}$Pelletron Linac Facility, Tata Institute of Fundamental Research, Mumbai - 400005, India}

\date{\today}
             
\begin{abstract}
To probe the role of the intrinsic structure of the projectile on sub-barrier fusion, measurement of fusion cross sections has been carried out in $^{9}$Be + $^{197}$Au system in the energy range E$_{c.m.}$/V$_B$ $\approx$ 0.82 to 1.16 using off-beam gamma counting method. Measured fusion excitation function has been analyzed in the framework of the coupled-channel approach using CCFULL code. It is observed that the coupled-channel calculations, including couplings to the inelastic state of the target and the first two states of the rotational band built on the ground state of the projectile, provide a very good description of the sub-barrier fusion data. At above barrier energies, the fusion cross section is found to be suppressed by $\approx$ 39(2)\% as compared to the coupled-channel prediction.  A comparison of reduced excitation function of $^{9}$Be + $^{197}$Au  with other $x$ + $^{197}$Au shows  a larger enhancement for $^9$Be in the sub-barrier region amongst Z=2-5 weakly and tightly bound projectiles, which indicates the prominent role of the projectile deformation in addition to the weak binding.

\end{abstract}

\maketitle

\section{\label{sec:level1}Introduction}
Nuclear reactions involving weakly bound stable ($^{6,7}$Li, $^{9}$Be) and unstable ($^{6,8}$He, $^{7,10,11}$Be) projectiles have been extensively investigated in recent years due to their importance in understanding the effect of coupling to the continuum and many-body quantum tunneling phenomenon~\cite{lf1,nk, lf2,bb}. In particular,  efforts have been made to understand the role of low break-up threshold of projectiles on the fusion cross sections arising from  extended shapes and $\alpha$ + x cluster structures. During the projectile-target interaction, a weakly bound projectile may break-up into the constituent $\alpha$-cluster(s) before reaching the fusion barrier. Hence, both the complete fusion (CF) - where the entire projectile fuses with the target nucleus, and break-up fusion or incomplete fusion (ICF) - where a part of the projectile fuses with the target nucleus, are observed. The projectile break-up results in the reduced incoming flux~\cite{nti, jt}, and therefore the cross section of CF is expected to be suppressed as compared to that for the tightly bound projectile. The observed sub-barrier fusion enhancement can be explained within the framework of the coupled-channel calculations by including couplings to the inelastic states and direct reaction channels such as neutron transfer and break-up.

Although the phenomenon of fusion suppression is widely accepted and attributed to the weak binding of nuclei, its origin is not yet fully understood. In reactions involving $^{6,7}$Li and $^{9}$Be projectiles with heavy mass targets, the complete fusion has been reported to be suppressed by $\approx$ 30$\%$ as compared to the standard coupled-channel calculations~\cite{lf1,nk,bb,rr,jla,md3, sg, tp, hgi, bw}.  
 A systematic study of the break-up effects on the complete fusion cross sections at energies above the Coulomb barrier is reported by Wang~\textit{et al.}~\cite{bw}. In this report, it has been shown that for a given projectile, the suppression effect is independent of the target. Generally, a strong correlation is observed between the suppression factor and the lowest break-up threshold energy~\cite{bw,amplb}. Hinde~\textit{et al.}~\cite{hinde2010} reported that in spite of widely different $\alpha$-break-up threshold energies, $^{9,10,11}$Be show a significant suppression of the complete fusion. Recently, Cook~\textit{et al.}~\cite{kjc} concluded that the cluster transfer rather than actual break-up prior to reaching the fusion barrier is responsible for fusion suppression.
 
 While the CF suppression with weakly bound nuclei has been studied in many systems, a comparative study of fusion cross-section enhancement at sub-barrier energies to probe the effect of the projectile structure have been sparse. Lemasson~\textit{et al.}~\cite{al} reported that complete fusion cross sections at sub-barrier energies for halo nucleus $^{8}$He were significantly enhanced as compared to $^{4}$He, mainly due to the coupling to the neutron transfer channel. Further, complete fusion cross section of $^{8}$He and $^{6}$He were found to be similar, which was attributed to the role of higher-order processes with neutron-pair transfer preceding fusion. 
 As mentioned earlier, the weakly bound stable $^{6}$Li, $^{7}$Li, $^{9}$Be nuclei are dominantly clusters of alpha-deuteron, alpha-triton, and alpha-alpha-neutron, respectively. In particular,  $^{9}$Be exhibits a Borromean structure ($\alpha$+$\alpha$+n) with a large deformation in the ground state. Moreover, the ground state is the only bound state of the system, and all excited states are particle unbound. Hence, reactions with $^9$Be at near barrier energies are important for a systematic study of weakly bound stable and unstable projectiles. 
 
 In the present work, the fusion cross sections in $^{9}$Be + $^{197}$Au system have been measured at near barrier energies and analyzed in the framework of the coupled-channel approach using theoretical model code CCFULL. The choice of $^{9}$Be~+~$^{197}$Au system is primarily driven by the fact that the fusion studies with different weakly bound projectiles, namely, $^{6}$He~\cite{yup}, $^{8}$He~\cite{al} and $^{6,7}$Li~\cite{cs} on $^{197}$Au target have been reported earlier. A comparative study of these systems, together with $^{11}$B~+~$^{197}$Au data ~\cite{asb},  enables the assessment of the impact of weak binding on sub-barrier fusion.
 
 This paper is organized as follows - experimental details are given in Section II, results and discussions of experimental data employing statistical model calculations and coupled-channel calculations are described in Section III.
 In Section IV, a systematic comparison of weakly bound projectiles on $^{197}$Au is presented, and a summary of the present work is given in Section V.

\section{\label{sec:level1}Experimental details}
The experiment was performed at the Pelletron Linac facility at TIFR, Mumbai, India. Self-supporting $^{197}$Au target foils of thickness $\sim$ 1.3 - 1.7~mg/cm$^{2}$ were prepared using the rolling technique. The $^{9}$Be beam in the energy range E$_{lab}$ $\approx$ 30 - 47 MeV was bombarded on the  $^{197}$Au target with a typical beam current of 8 - 15 pnA. The Aluminium catcher foils of the appropriate thickness ($\sim$ 1.5~mg/cm$^{2}$) were mounted behind the target foils to stop the recoiling reaction residues. For the effective utilization of beamtime, some of the irradiations were performed using a stack of two target-catcher foil assemblies.  The incident energy and the energy spread at half target thickness were calculated using SRIM~\cite{tr}. In order to correct for beam fluctuations during the irradiation, the beam current was recorded at regular intervals of 30 or 60 seconds using a CAMAC (Computer Automated Measurement And Control) scaler. The gamma-rays from irradiated samples were counted off-line using two efficiency calibrated HPGe detectors. The energy calibration and efficiency measurement of the HPGe detectors were carried out using a  standard precalibrated $^{152}$Eu $\gamma$- ray source. Both the HPGe detectors were shielded with $\sim$5 cm thick lead rings for reducing the ambient background. Data were recorded using a digital data acquisition system (DAQ) employing CAEN digitizer (14 bit ADC, 100 MHz sampling rate), and the off-line data analysis were performed using LAMPS software~\cite{lam}.  Typical energy resolution obtained is about 2.4~keV at 1408~keV. The off-line counting was performed either at a distance of 10 cm from the face of the detector or in the close geometry (in which the sample was mounted on the face of the detector) depending on the activity of the irradiated sample. For the two lowest energies, target and catcher foils were counted separately to improve the sensitivity. 

\begin{table}[h!]
\caption{Evaporation residues (ER) from complete fusion in $^9$Be~+~$^{197}$Au reaction together with  half-life (T$_{1/2}$), and  energy (E{$_\gamma$}) and absolute intensity (I$_{\gamma}$) of prominent gamma-rays~\cite{nndc}.}
\begin{ruledtabular}
\begin{tabular}{lllll} 
 Channel & ER & T$_{1/2}$  & E$_{\gamma}$ (keV) & I$_{\gamma}$($\%$)\\ [0.5ex] 
 \hline
 2n & $^{204}$Bi &  11.22 h  &  374.7    & 82 \\
          &             &         &  984   & 59    \\
          &             &         &  899.15    & 99    \\
 3n & $^{203}$Bi &  11.76 h  & 820.2    &30 \\
    &             &         &  825.2    & 14.8    \\
       &             &         &  896.9    & 13.2    \\
 4n & $^{202}$Bi & 1.71 h  &   422.13    & 83.7 \\
      &             &         &  657.49    & 60.6    \\
      &             &         &  960.67    &  99.3    \\
  5n & $^{201}$Bi &  1.72 h &  629.1  & 26 \\
       &             &         &  936.2    & 12.2    \\
      &             &         &  1014.1    & 11.6   \\
\end{tabular}
\label{table1}
\end{ruledtabular}
\end{table}

\begin{figure}

\vspace{0cm}
   \includegraphics[trim=1.5cm 1.2cm 1cm 1.2cm, width=8cm]{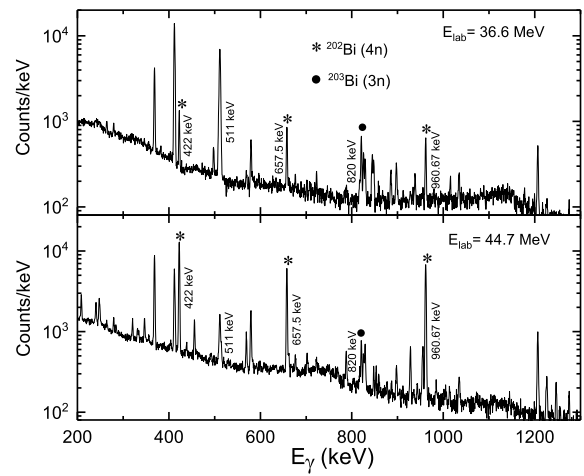}
\caption{Off-line $\gamma$ ray spectra measured in $^{9}$Be + $^{197}$Au reaction at E$_{lab}$ = 36.6 and 44.7~MeV. The characteristic gamma-rays of ERs $^{203}$Bi, and $^{202}$Bi are marked.}
\label{fig1}

  \vspace{1cm}
\end{figure}

  Table~\ref{table1} gives a summary of expected ERs and their characteristics gamma-rays.  Fig.~\ref{fig1} shows off-line $\gamma$-ray spectra obtained for $^{9}$Be~+~$^{197}$Au reaction at E$_{lab}$ = 36.6  and 44.7 MeV. The ERs $^{203,202}$Bi have been identified by their characteristic $\gamma$-rays, which are marked in the figure. The identification of gamma-rays was confirmed by half-life measurement and verification of relative yields of multiple gamma-rays of a given residue.
  From the observed photopeak yield  N$_\gamma$, the ER cross section ($\sigma_{x}$) can be calculated as,
 \begin{equation}
 \sigma_{x} = \frac{N_{\gamma} \lambda_{x} t_{irr}}{I_{\gamma} \epsilon_{\gamma} ({e^{-\lambda_{x}t_{1}}-e^{-\lambda_{x}t_{2}}}) N_{P} N_{T} ({1-e^{-\lambda_{x}t_{irr}}}) }
 \end{equation}
where,  N$_{P}$ is the number of incident particles, N$_{T}$ is the number of target particles per unit area, $\lambda_{x}$ is a decay constant,  t$_{irr}$ is the duration of irradiation,  t$_{1}$ (t$_{2}$) is the time since the end of irradiation to the start (end) of counting, I$_{\gamma}$ and $\epsilon_{\gamma}$ are  the absolute intensity  and the photopeak efficiency of the characteristic $\gamma$-ray, respectively. This equation takes into account the decay during the irradiation (t$_{irr}$) and  assumes the uniform beam current. As mentioned earlier, the beam current was recorded in smaller intervals to take care of the fluctuations, and the decay corrections~\cite{ntz} were applied to each interval for computation of $\sigma_{x}$. From the recorded beam charge $Q$,  N$_{P}=Q/q_{eq}$ is calculated, where $q_{eq}$ is the equilibrium charge state. The value of q$_{eq}$ is found to be +4 from the theoretical calculation~\cite{gs} and from the prediction of code CHARGE of LISE++~\cite{lsc}, over the range of energy and target assembly thicknesses studied in the present work.
  
\section{\label{sec:level1}Results and Discussions}

\begin{table*}
\caption{ Measured ER cross sections in $^{9}$Be + $^{197}$Au  reaction (V$_{B}$ = 38.4 MeV). }
\begin{ruledtabular}
\begin{tabular}{llllll}
E$_{lab}$ (MeV) & E$_{c.m.}$ (MeV) & $^{202}$Bi (mb) & $^{203}$Bi (mb) &  R  & $\sigma^{Corr}_{CF}$(mb)\\ \hline
 33    & 31.6 & - & 0.020 $\pm$ 0.004 & 0.96 &  0.021 $\pm$ 0.004  \\ 
 
  34.5  & 33 & - & 0.20 $\pm$ 0.01 & 0.98  & 0.20 $\pm$ 0.01  \\ 
 
    35.6  & 34 & 0.10 $\pm$ 0.03 & 1.1 $\pm$ 0.1  &  0.99 &  1.2 $\pm$ 0.1 \\ 
    
      36.6  & 35 & 0.70 $\pm$ 0.04 & 4.5 $\pm$ 0.4  &  0.99 &  5.3 $\pm$ 0.4 \\
      
         37.6  & 36 & 3.7 $\pm$ 0.2 & 12 $\pm$ 1  &  0.99 &   16 $\pm$ 1\\ 
      
           38.6  & 36.9 & 12.7 $\pm$ 0.1 & 21 $\pm$ 1  & 0.99 &  34 $\pm$ 1 \\ 
           
           39.6  & 37.9 & 31.4 $\pm$ 0.2 & 32 $\pm$ 2 & 0.99  &   64 $\pm$ 2 \\

     39.9  & 38.2 & 54.5 $\pm$ 0.3 & 32 $\pm$ 3 & 0.99 &  87 $\pm$ 3  \\
    
         40.6  & 38.8 & 69 $\pm$ 1 & 40 $\pm$ 2 & 0.99 &  111 $\pm$ 2  \\
    
      42.7  & 40.8 & 155 $\pm$ 8 & 35 $\pm$ 3 & 0.99 &  192 $\pm$ 9 \\

     44.7  & 42.7 & 281 $\pm$ 3 & 32 $\pm$ 5 & 0.98  &  320 $\pm$ 6\\

46.7  & 44.7 & 328 $\pm$ 6 & 24 $\pm$ 3 & 0.96  &  367 $\pm$ 7  \\ 
    
\end{tabular}
\end{ruledtabular}
\label{csdata}
\end{table*}

\begin{figure}[h]
    \includegraphics[trim=1cm 2cm 0.5cm 2cm, width=8.5cm]{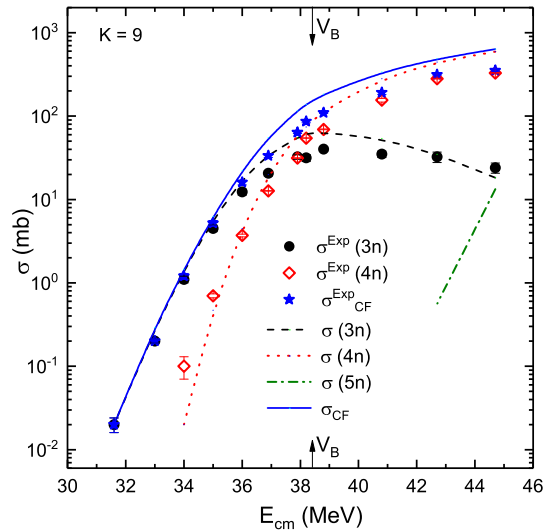}
    \caption{(Color online) Measured excitation functions of evaporation residues $^{202}$Bi (4n) and $^{203}$Bi (3n)  in $^{9}$Be + $^{197}$Au system together with the statistical model calculations. The fusion cross section $\sigma_{CF}$ (obtained from CCFULL) and $\sigma^{Exp}_{CF}$=($\sigma^{Exp}_{3n+4n}$) are also shown for comparison. The barrier  obtained from the CCFULL calculations, V$_{B}$ = 38.4 MeV, is indicated for reference.}
    \label{csplot}
\end{figure}

  In the present work, the fusion cross sections are measured down to 18$\%$ below the Coulomb barrier. Measured cross sections of $^{203}$Bi(3n) and $^{202}$Bi(4n) are listed in Table~\ref{csdata}. Errors shown in the cross sections are statistical. It may be noted that the contribution of evaporation residues $^{204}$Bi (2n) and $^{201}$Bi (5n), and the fission are expected to be small and could not be unambiguously measured at the present level of sensitivity. The measured ER cross sections are compared with PACE2~\cite{ag} calculations. The PACE2 is a statistical model code, which employs Monte Carlo simulations to calculate the decay of the compound nucleus using the Hauser-Feshbach approach. In the present calculation, Ignatyuk level density prescription~\cite{igk} was used with an asymptotic level density parameter a=A/K, where K is varied between 8-10.
At all energies, the angular momentum (${\ell}$) distribution (and hence $\sigma_{CF}$) obtained from the CCFULL calculations (inclusive of couplings - as described later)  has been used as input in the PACE2. A comparison of experimentally measured and theoretically calculated excitation functions of evaporation residues $^{202}$Bi (4n) and $^{203}$Bi (3n)  in $^{9}$Be + $^{197}$Au system is presented in Fig.~\ref{csplot}. It is observed that $a$ = A/9 MeV$^{-1}$ gives the best agreement with the experimental data.

The statistical model calculations show that neutron evaporation channels are dominant, and exhaust $\approx$99\% of CF cross section over most of the experimentally measured energies, consistent with other systems in this mass range~\cite{st, ast}. As mentioned earlier, $\sigma_{5n}$, $\sigma_{2n}$ and fission could not be measured. Therefore, the contribution of missing channels has been deduced~\cite{vv} using  PACE2. The ratio of xn channels (3n,4n) and the complete fusion cross section,  R=($\sigma_{3n+4n}^{PACE}$)/$\sigma_{fus}$ is determined using the PACE2 calculations at different energies and the experimental fusion cross section is derived as $\sigma^{Corr}_{CF}$=$\sigma_{3n+4n}^{exp}$/R. The values of R and corrected cross sections ($\sigma^{Corr}_{CF}$) are given in Table~\ref{csdata}.  
At higher energies, the correction mainly arises due to missing $\sigma_{5n}$ and fission, while that at lower energies it is due to missing $\sigma_{2n}$. It is important to note that the maximum correction is $\sim4\%$ at the extreme energy points and hence has no significant effect on the conclusions drawn in the present work. 

\begin{figure}

    \includegraphics[trim=1cm 4.5cm 1cm 1cm, width = 8cm]{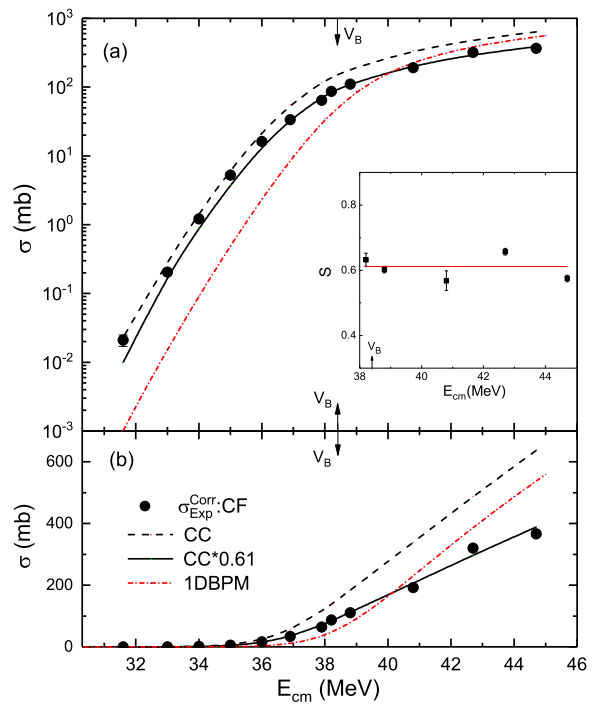}
    \caption{(Color online)  (a) The measured CF excitation function of $^{9}$Be + $^{197}$Au system  together with  CCFULL calculations, (b)  data in panel (a) on a linear scale for better visualization of the experimental data and theoretical calculations at the above-barrier energies. Inset in panel(a) shows the suppression factor S at above-barrier energies (see text for details).}
    \label{fig:ccfull}
\end{figure}

 At sub-barrier energies, the fusion cross sections are calculated with the CCFULL code modified specially for $^{9}$Be projectile~\cite{kh,md}. The CCFULL calculations, without incorporating couplings to any inelastic excitation, provide a simple one-dimensional barrier penetration model (1DBPM) for easy reference. The coupled-channel calculations performed using CCFULL are presented in Fig.~\ref{fig:ccfull}. The potential parameters used in the calculations, namely, V$_{0}$ = 51.94 MeV, r$_{0}$ = 1.17 fm, a$_{0}$ = 0.63 fm, are taken from the Woods-Saxon parametrization of the Akyuz-Winther (AW) potential~\cite{rab}. The calculations  include the coupling of both projectile and target excited states. For $^{9}$Be, the ground state spin $\frac{3}{2}^{-}$ with the deformation parameter $\beta_{2}$ = 1.3~\cite{hjv} and the first  two excited states in K = $\frac{3}{2}^{-}$ (band head) ground-state rotational band ~\cite{hn} are taken into consideration. In case of $^{197}$Au, the inelastic excitation to the first excited state at E$_{x}$ = 0.077 MeV is included as a vibrational state with $\beta$ = 0.1~\cite{xh}. It is evident from Fig.~\ref{fig:ccfull} that CCFULL output shows good agreement with data at sub-barrier energies, but over-predicts the data at near- and above-barrier energies.
It should be mentioned that for $^9$Be projectile, different $\beta_{2}$ values have been used in CCFULL calculations. For example, $\beta_{2}$ = 0.92 in Ref.~\cite{md},  the best fit value of Ref.~\cite{hjv} $\beta_{2}$ = 1.1, and $\beta_{2}$ = 1.3~\cite{csbe,vivekbe}. In the present case, $\beta_{2}$ = 1.3 was found to describe the data well with the above mentioned AW potential.  

The suppression factor S~=~$\sigma^{Corr}_{CF}$/$\sigma^{CC}_{CF}$  for energies above the barrier is shown in the inset of Fig.~\ref{fig:ccfull}(a). The mean value of S=0.61(2), implies $\approx39\pm2$\% CF suppression for $^{9}$Be + $^{197}$Au system due to the involvement of the weakly bound projectile.  The CCFULL output scaled with S=0.61 is also shown in the same figure (solid line), which matches reasonably well with the measured excitation function.  The observed suppression factor is consistent with 25-40$\%$ fusion suppression observed with $^{9}$Be on different heavy targets~\cite{lf1,lf2, md2, yd1}.

\section{\label{sec:level1} Comparison of weakly bound projectiles: $\rm x$ + $\rm ^{197}Au$ systems}
To understand the role of projectile structure on fusion involving weakly bound nuclei, where break-up is a dominant channel, a systematic comparison of fusion excitation functions have been carried out for different $x$ + $^{197}$Au combinations. The values of break-up threshold energy for various weakly bound projectiles,  calculated using the latest mass tables,  are tabulated in Table~\ref{sep_energy}.
 
\begin{table}[h!]
\caption{List of dominant break-up channels together with corresponding break-up threshold energy ($E_{BU}$) for weakly bound projectiles considered in the present study.}

\begin{ruledtabular}
\begin{tabular}{lll} 
 Nuclei & Channel & E$_{BU}$ (MeV) \\ [0.5ex] 
 \hline
  $^{9}$Be & $\alpha$~+~$\alpha$~+~n & 1.575 \\ 
           & $^{8}$Be~+~n  & 1.667 \\
 $^{7}$Li &  $\alpha$~+~t  & 2.467\\
  $^{6}$Li & $\alpha$~+~d  & 1.473\\
   $^{8}$He &  $^{6}$He~+~2n &  2.125\\
           &  $^{7}$He~+~n  & 2.535  \\
  $^{6}$He &$\alpha$~+~2n &  0.975\\
\end{tabular}
\end{ruledtabular}
\label{sep_energy}
\end{table}
 
For comparison of different projectile-target systems, appropriate scaling of cross sections is essential. Scaling methodologies have been extensively discussed in Ref.~\cite{lfc2015}. Canto \textit{et al.}~\cite{lf} introduced the reduced fusion cross section and the reduced energy variables, defined as,
\begin{equation}
    E_{red} = \frac{E_{c.m.}-V_{B}}{\hbar\omega} ~~~{\rm and} ~~~
    \sigma_{red} = \frac{2.E_{c.m.}}{\hbar\omega.R_{B}^{2}}  \sigma_{F}\\
\end{equation}
 
 \begin{figure*}

    \includegraphics[trim=9cm 2cm 9cm 1cm, width = 7.5cm]{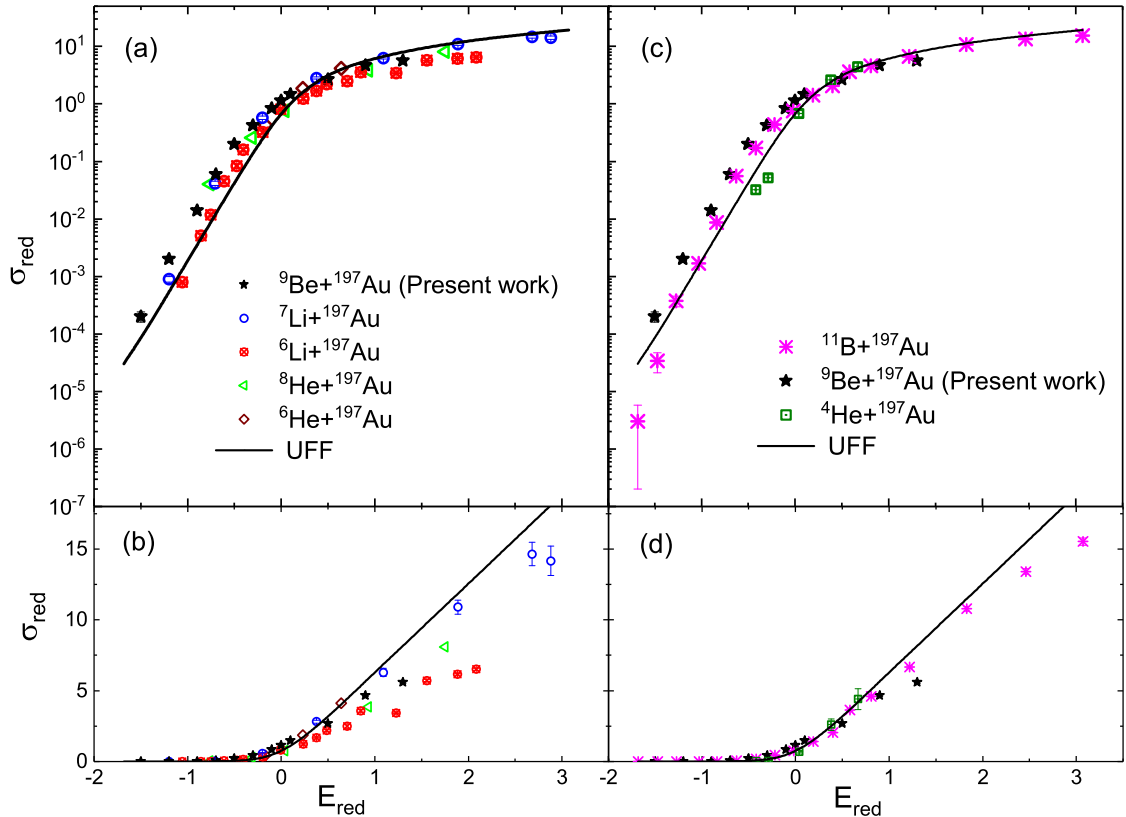}
       \caption{(color online)
      A comparison of reduced fusion excitation functions of x + $^{197}$Au: (a) weakly bound projectiles $^{6}$He~\cite{yup}, $^{8}$He~\cite{al}, $^{6,7}$Li~\cite{cs}, and (c) projectiles with higher break-up threshold $^{4}$He~\cite{msb}, $^{11}$B~\cite{asb}. Panels (b) and (d) show data of (a) and (c), respectively,  on a linear scale for better visibility of the above-barrier region.  The UFF as defined in eq.(3) is also plotted for reference (see text for details).
       }
       
    \label{ff_red}

\end{figure*}

It is evident that these reduced variables depend on the radius and height of the fusion barrier, and the barrier curvature ($\hbar\omega$), thereby taking into consideration static as well as dynamic effects. The available data for weakly bound projectiles on $^{197}$Au has been analyzed using the above scaling method. In addition, data with projectiles having higher break-up threshold energy, namely, $^4$He (E$_{BU}$=20.578 MeV, for n+$^3$He) and  $^{11}$B ((E$_{BU}$= 8.664 MeV, for $\alpha$+$^{7}$Li) is also analyzed in the same framework.  It should be mentioned that for both $^6$He and $^4$He, data points with large error bars are not considered in the present analysis, and only a subset of reported data is used \cite{al}. The barrier parameters used for obtaining reduced variables are listed in Table~\ref{redpar}.  These parameters are obtained from the CCFULL calculations for $^{6,7}$Li~\cite{cs}, $^9$Be~(present data), and $^{11}$B~\cite{asb}. Those for $^{4,6,8}$He are derived from the fusion cross section data and scaled data presented in Ref.~\cite{al}. 
\begin{table}[h!]
\caption{Barrier parameters: V$_{B}$ (MeV), R$_{B}$ (fm), $\hbar\omega$~(MeV) used to obtain $\sigma_{red}$ and $E_{red}$}

\begin{ruledtabular}
\begin{tabular}{lllll} 
  System & V$_{B}$ (MeV) & R$_{B}$ (fm) &  $\hbar\omega$ (MeV) \\ [0.5ex] 
 \hline
 $^{11}$B~+~$^{197}$Au & 46.87  & 11.42 & 4.65  \\ 
  $^{9}$Be~+~$^{197}$Au & 38.4  & 11.14 & 4.72  \\ 
  $^{6}$Li~+~$^{197}$Au & 28.92  & 11.13 & 5.13\\
  $^{7}$Li~+~$^{197}$Au & 29.28  & 11.00 & 4.86\\
  $^{8}$He~+~$^{197}$Au & 18.65 & 11.60  &  3.47\\
  $^{6}$He~+~$^{197}$Au & 19.13  & 11.12 & 4.24\\
  $^{4}$He~+~$^{197}$Au & 19.77  & 10.79  & 5.39\\
  
\end{tabular}
\end{ruledtabular}
\label{redpar}
\end{table}
 Fig.~\ref{ff_red}(a) shows the reduced fusion excitation functions of different weakly bound projectiles, namely, $^{6,8}$He~\cite{al,yup}, $^{6,7}$Li~\cite{cs} and $^{9}$Be (present work), on $^{197}$Au target. The same data is also shown on a linear scale in Fig.~\ref{ff_red}(b)  for the better visualization of data at above-barrier energies. 
A comparison of reduced excitation function of weakly bound $^{9}$Be with $^4$He and $^{11}$B (projectiles having higher break-up threshold energy)
is shown in Fig.~\ref{ff_red}(c) and (d) on a logarithmic and linear scale, respectively. The Universal Fusion Function (UFF), which represents Wong's formula in the absence of coupling effects, given by
 \begin{equation}
     F_{0}(x) = ln[1+exp(2\pi x)]
 \end{equation}
is also plotted in the same figure for reference. All weakly bound projectiles exhibit the expected common feature, namely,  enhancement below the barrier and suppression above the barrier (w.r.t UFF). However, it should be noted that the reduced variables are sensitive to R$_b$ and $\hbar\omega$ and hence are model dependent. It is interesting to see that $^9$Be shows the highest sub-barrier fusion enhancement in this group of Z=2-5 projectiles, which include halo nuclei $^{6,8}$He. Even amongst the stable weakly bound nuclei, $\sigma_{red}$ of $^{9}$Be is $\sim$ factor of 2 higher as compared to  $^{6}$Li, which has similar E$_{BU}$. Although couplings to transfer and break-up channels are not included in coupled-channel calculations for $^9$Be, the CCFULL calculations including couplings to rotational states can explain the present data at sub-barrier energies (see Fig.~\ref{fig:ccfull}). Thus the observed large sub-barrier enhancement points to the important role of deformation, even for weakly bound nuclei. The fusion suppression factor measured in the present work for  $^{9}$Be~+~$^{197}$Au system (39$\pm2\%$) is similar to that for $^{6}$Li+$^{197}$Au~\cite{cs} (35$\pm2\%$). Although data for fusion suppression $^{6,8}$He~+~$^{197}$Au is not reported, it can be seen from Fig.~\ref{ff_red}(b) that at above-barrier energies CF suppression in case $^{8}$He is less compared to $^{9}$Be and $^6$Li, which is consistent with break-up threshold energies.   

\section{\label{sec:level1}Summary and Conclusions}
The fusion excitation function in ${^9}$Be + $^{197}$Au reaction has been measured in the energy range 0.82 $\leq$ E$_{c.m.}$/V$_{B}$ $\leq$ 1.16. The cross sections of evaporation residues $^{203}$Bi(3n) and $^{203}$Bi(4n) have been obtained by off-line gamma counting. The measured fusion excitation function is analyzed using theoretical model code CCFULL. The coupled-channel code CCFULL, including coupling to inelastic excitations of projectile and target nuclei, provides an excellent description of the experimental fusion excitation function at sub-barrier energies but under-predicts the same at above barrier energies. The experimental fusion cross section above the barrier is found to be suppressed by  $\approx$39(2)\%  as compared to the coupled-channel calculations. In order to investigate the role of the intrinsic structure of projectile on sub-barrier fusion cross section, a systematic comparison of the reduced excitation functions of different projectile systems $x$ + $^{197}$Au is presented. It is observed that $\sigma_{red}$($^{9}$Be) shows a larger enhancement in the sub-barrier region amongst all Z=2-5 projectiles, namely, $^{4,6,8}$He, $^{6,7}$Li, and $^{11}$B,   which indicates the prominent role of the projectile deformation in addition to the weak binding.

\section{\label{sec:level1}Acknowledgments}
The authors would like to thank the PLF staff for providing the steady and smooth beam during the experiments, and the target lab personnel for their help in the target preparation.


\begin{thebibliography}{}
\bibitem{lf1} L. F. Canto \textit{et al.}, Phys. Rep. {\bf 424}, 1 (2006).
\bibitem{nk} N. Keeley \textit{et al.}, Prog. Part. Nucl. Phys. {\bf 59}, 579 (2007).
\bibitem{bb} B. B. Back \textit{et al.}, Rev. Mod. Phys. {\bf 86}, 317 (2014).
\bibitem{lf2} L. F. Canto \textit{et al.}, Phys. Rep. {\bf 596}, 1 (2015) and references therein.
\bibitem{nti} N. Takigawa \textit{et al.}, Phys. Rev. C {\bf 47}, R2470 (1993).
\bibitem{jt} J. Takahashi \textit{et al.}, Phys. Rev. Lett. {\bf 78}, 30 (1997).
\bibitem{rr} R. Rafiei \textit{et al.}, Phys. Rev. C {\bf 81}, 024601 (2010).
\bibitem{jla} Jin Lei and Antonio M. Moro, Phys. Rev. Lett. {\bf 122}, 042503 (2019).
\bibitem{md3} M. Dasgupta \textit{et al.}, Phys. Rev. Lett. {\bf 82}, 1395 (1999).
\bibitem{sg} C. Signorini \textit{et al.}, Eur. Phys. J. A {\bf 10}, 249 (2001).
\bibitem{tp} V. Tripathi \textit{et al.}, Phys. Rev. Lett. {\bf 88}, 172701 (2002).
\bibitem{hgi} K. Hagino \textit{et al.}, Phys. Rev. C {\bf 61}, 037602 (2000).
\bibitem{bw} Bing Wang \textit{et al.}, Phys. Rev. C {\bf 90}, 034612 (2014).
\bibitem{amplb} A. Mukherjee \textit{et al.}, Phys. Lett. B {\bf 636}, 91 (2006).
\bibitem{hinde2010} D.J. Hinde \textit{et al.} Phys. Rev. C {\bf 81}, 064611 (2010).
\bibitem{kjc} K. J. Cook \textit{et al.}, Phys. Rev. Lett. {\bf 122}, 102501 (2019).
\bibitem{al} A. Lemasson \textit{et al.}, Phys. Rev. Lett. {\bf 103}, 232701 (2009).
\bibitem{yup} Yu. E. Penionzhkevich \textit{et al.}, Eur. Phys. J. A {\bf 31}, 185 (2007).
\bibitem{cs} C. S. Palshetkar \textit{et al.}, Phys. Rev. C {\bf 89}, 024607 (2014).
\bibitem{asb} A. Shrivastava \textit{et al.}, Phys. Rev. C {\bf 96}, 034620 (2017).
\bibitem{tr} http://www.srim.org
\bibitem{lam} http://www.tifr.res.in/$\sim$pell/lamp.html
\bibitem{nndc} https://www.nndc.bnl.gov/
\bibitem{ntz} N. T. Zhang et. al., Phys. Rev. C {\bf 90}, 024621 (2014). 
\bibitem{gs} G. Schiwietz \textit{et al.}, Nucl. Instr. and Meth. B {\bf 175}, 125 (2001).
\bibitem{lsc} http://lise.nscl.msu.edu
\bibitem{ag} A. Gavron, Phys. Rev. C {\bf 21}, 230 (1980).
\bibitem{igk} A.V. Ignatyuk \textit{et al.}, Sov. J. Nucl. Phys. {\bf 21}, 255 (1975).
\bibitem{st} Shital Thakur et. al., Euro. Phys. J. Web of Conferences {\bf 17}, 16017 (2011).
\bibitem{ast} A. Shrivastava et. al., Phys. Lett. B {\bf 718}, 931 (2013).
\bibitem{vv} V. V. Parkar \textit{et al.}, Phys. Rev. C {\bf 98}, 014601 (2018).
\bibitem{kh} K. Hagino \textit{et al.}, Phys. Commun. {\bf 123}, 143 (1999).
\bibitem{md} M. Dasgupta \textit{et al.}, Phys. Rev. C {\bf 70}, 024606 (2004).
\bibitem{rab} R. A. Broglia and A. Winther, Elastic and Inelastic Reactions, Heavy Ion Reactions, Heavy Ion Reaction Lecture Notes Vol. I (Benjamin Cummings Redwood City, CA, 1981), p. 114.
\bibitem{hjv} H. J. Votava \textit{et al.}, Nucl. Phys. A {\bf 204}, 529 (1973).
\bibitem{hn} H. Nguyen Ngoc \textit{et al.}, Nucl. Phys. {\bf 42}, 62 (1963).
\bibitem{xh} Xiaolong Huang and Chunmei Zhou, Nucl. Data Sheets {\bf 104}, 283 (2005).
\bibitem{csbe} C.S. Palshetkar \textit{et al.}, Phys. Rev. C {\bf 82}, 044608 (2010).
\bibitem{vivekbe} V.V. Parkar \textit{et al.}, Phys. Rev. C {\bf 82}, 054601 (2010).
\bibitem{md2} M. Dasgupta \textit{et al.}, Phys. Rev. C {\bf 81}, 024608 (2010).
\bibitem{yd1} Y. D. Fang \textit{et al.}, Phys. Rev. C {\bf 91}, 014608 (2015).
 \bibitem{lfc2015} L.F. Canto \textit{et al.}, Phys. Rev. C {\bf 92}, 014626 (2015).
 \bibitem{lf}L. F. Canto \textit{et al.}, J. Phys. G. Nucl. Part. Phys. {\bf 36} 015109 (2009), Nucl. Phys. A {\bf 821}, 51 (2009).
\bibitem{msb} M.S. Basunia \textit{et al.}, Phys. Rev. C {\bf 75}, 015802 (2007).

\end{thebibliography}
\end{document}